%% file: main.tex
\documentclass{article} % For LaTeX2e
\usepackage{iclr2023_conference,times}
\usepackage{microtype}

\usepackage{amsmath}
\usepackage{amssymb}
\usepackage{mathtools}
\usepackage{amsthm}
\usepackage{graphicx}
\usepackage{booktabs}
\usepackage{colortbl}
\usepackage{amsmath}
\usepackage{amssymb}
\usepackage{amsfonts}
\usepackage{multirow}
\usepackage{verbatim}
\usepackage{caption}
\usepackage{longtable}
\usepackage{supertabular}
\usepackage{float}

\usepackage{xcolor}
\usepackage{xspace}
\usepackage{textcomp}
\usepackage{makecell}
\usepackage{multirow}
\usepackage{multicol}
\usepackage{lscape} 
\usepackage{siunitx}
\usepackage{algorithm}
\usepackage{algpseudocode}

\usepackage{amssymb}% http://ctan.org/pkg/amssymb
\usepackage{pifont}% http://ctan.org/pkg/pifont
\usepackage{scrextend}

\usepackage{array}
\usepackage{latexsym}
\usepackage[T1]{fontenc}
\usepackage[utf8]{inputenc}
\usepackage{microtype}
\usepackage{url}            % simple URL typesetting
\usepackage{nicefrac}       % compact symbols for 1/2, etc.
\usepackage{changepage}
\usepackage{xargs}
\usepackage{wrapfig,lipsum,booktabs}
\usepackage{longtable}
\usepackage{subcaption}
\usepackage{pgfplots}
\usetikzlibrary{pgfplots.groupplots}
\pgfplotsset{compat=1.3}
\usepackage{tikz}
\usetikzlibrary{patterns}

\usepackage[most]{tcolorbox}

\usepackage{hyperref}
\usepackage[capitalize,noabbrev]{cleveref}
\crefname{section}{Section}{\S\S}
\crefname{section}{Section}{\S\S}
\crefname{table}{Table}{Tables}
\crefname{figure}{Figure}{Figures}
\crefname{algorithm}{Algorithm}{}
\crefname{equation}{eq.}{}
\crefname{appendix}{Appendix}{}
\crefformat{section}{Section #2#1#3}
\crefformat{footnote}{#2\footnotemark[#1]#3}

\definecolor{mydarkblue}{rgb}{0,0.08,0.45}
\hypersetup{
    colorlinks=true,
    linkcolor=mydarkblue,
    filecolor=blue,      
    urlcolor=mydarkblue,
    citecolor=mydarkblue,
}

\input{math_commands}

\usepackage{wrapfig}
\usepackage[font=small,skip=1pt]{caption}
\usepackage{balance}
\usepackage{paralist}
\usepackage{subcaption}
\usepackage{cleveref}
\usepackage{colortbl}
\usepackage{algorithm}
\usepackage{algpseudocode}
\usepackage[nolist]{acronym}
\usepackage{tabularx}
\usepackage{makecell}
\usepackage{booktabs}
\definecolor{OliveGreen}{rgb}{0,0.4,0}

\usepackage[symbol]{footmisc}

\usepackage{authblk}
\usepackage{relsize}
\iclrfinalcopy

\title{Better Generalization with Semantic IDs:\\ A Case Study in Ranking for Recommendations}

\author[*1]{Anima Singh}
\author[*2]{Trung Vu}
\author[*1]{Nikhil Mehta}
\author[2]{Raghunandan Keshavan}
\author[1]{Maheswaran Sathiamoorthy}
\author[2]{Yilin Zheng}
\author[1]{Lichan Hong}
\author[2]{Lukasz Heldt}
\author[2]{Li Wei}
\author[2]{Devansh Tandon}
\author[1]{Ed H. Chi}
\author[1]{Xinyang Yi}
\affil[1]{Google DeepMind}
\affil[2]{Google}

\begin{document}
\maketitle
\footnotetext{\textsuperscript{*}Equal contributions. Correspondence to \{\texttt{animasingh}, \texttt{trungtvu}, \texttt{nikhilmehta}\}@google.com}

\begin{abstract}
Randomly-hashed item ids are used ubiquitously in recommendation models. However, the learned representations from random hashing prevents generalization across similar items, causing problems of learning unseen and long-tail items, especially when item corpus is large, power-law distributed, and evolving dynamically. In this paper, we propose using content-derived features as a replacement for random ids. We show that simply replacing ID features with content-based embeddings can cause a drop in quality due to reduced memorization capability. To strike a good balance of memorization and generalization, we propose to use Semantic IDs \cite{rajput2023recommender} -- a compact discrete item representation learned from frozen content embeddings using RQ-VAE that captures the hierarchy of concepts in items -- as a replacement for random item ids. Similar to content embeddings, the compactness of Semantic IDs poses a problem of easy adaption in recommendation models. We propose novel methods for adapting Semantic IDs in industry-scale ranking models, through hashing sub-pieces of of the Semantic-ID sequences. In particular, we find that the SentencePiece model~\cite{kudo2018subword} that is commonly used in LLM tokenization outperforms manually crafted pieces such as N-grams. To the end, we evaluate our approaches in a real-world ranking model for YouTube recommendations. Our experiments demonstrate that Semantic IDs can replace the direct use of video IDs by improving the generalization ability on new and long-tail item slices without sacrificing overall model quality.
\end{abstract}

\section{Introduction}
\label{sec:1_introduction}
\input{1_introduction}

\section{Related Work}
\label{sec:2_related_work}
\input{2_related_work}

\section{Proposed Approaches}
\label{sec:4_proposed_approach}
\input{4_proposed_approach}
\section{Experiments}
\label{sec:5_experiments}
\input{5_experiments}

\section{Conclusion and Future Work}
\label{sec:6_conclusion}
\input{6_conclusion}

\bibliographystyle{abbrvnat}
\bibliography{references}

\clearpage
\appendix
\input{appendix}

\end{document}

%% file: math_commands.tex
\usepackage{amsmath,amsfonts,bm}
\usepackage{bbm}
\usepackage[mathscr]{eucal}

\def\1{\bm{1}}
\def\ve{{\bm{e}}}

\def\vr{{\bm{r}}}

\def\vx{{\bm{x}}}

\def\vz{{\bm{z}}}

\DeclareMathAlphabet{\mathsfit}{\encodingdefault}{\sfdefault}{m}{sl}
\SetMathAlphabet{\mathsfit}{bold}{\encodingdefault}{\sfdefault}{bx}{n}

\def\cC{{\mathcal{C}}}
\def\cD{{\mathcal{D}}}
\def\cE{{\mathcal{E}}}

\def\cL{{\mathcal{L}}}

\newcommand{\set}[1]{\{#1\}}

%% file: 1_introduction.tex
Neural models with large embedding tables are widely used in industry-scale recommender systems for scoring and ranking vast collections of items. These tables, often containing millions or even billions of rows, facilitate rapid memorization of item quality by modeling randomly-hashed item identifiers. It's worth noting that learning good item representations is crucial for personalization, as users are typically modeled as a sequence of items. Concretely, in this paper, we consider a neural ranking in a video recommendation system at YouTube. In this model, every video gets a unique identifier referred to as video ID, which is a random string devoid of meaning. This approach is widely adopted in numerous industry-scale recommender systems (e.g., \cite{cheng2016wide, kim2007music, koren2009matrix, zhao2019recommending}).

%Recommender systems are widely used across the industry to serve personalized content to users. They play a critical role in helping users discover novel content such as apps~\cite{cheng2016wide}, music~\cite{kim2007music, koren2009matrix}, and videos~\cite{covington2016deep, zhao2019recommending, gomez2015netflix}. This paper considers a neural ranking model in a large industrial-scale video recommendation system at YouTube. Our recommendation corpus has billions of videos. Every video gets a unique identifier (called video ID), a random string devoid of meaning. The ranking model gets as input multiple video IDs, for example, what the user is watching and what to rank. Furthermore, users' features are typically represented as a list of video IDs they have previously watched - a critical signal for personalization. Given that the central goal of a recommendation system is to connect users to videos, learning good representations of video ID is critical. 

In this paper, we study content-based item representations that can improve the generalization for new and long-tail item distributions while keeping models' power of memorization without sacrificing overall quality, with a focus on recommendation ranking models. A common technique for encoding item id is to learn one-hot embeddings. However, given an extremely large item corpus with billions of videos, learning one embedding vector per video can be 
resource-intensive, and more importantly, are vulnerable to the data sparsity of torso and tail items. For using a limited number of embeddings, an alternative approach is to use the hashing trick~\cite{weinberger2009feature} that maps many items to the same row. This approach can cause random collisions when the original item IDs are not semantically meaningful. When it comes to using content embeddings from pre-trained multimodal item encoders, it is unclear if large item ID table can be fully replaced due to the loss of item-level memorization. In ~\cite{yuan2023recommender}, authors show that frozen item embeddings outperformed item ID baselines for SASRec~\cite{sasrec}, but not for two-tower models~\cite{rendle2020neural} for datasets with up to $150k$-size corpus. In our experiments in YouTube with a much larger corpus, we observed a significant quality reduction (Section~\ref{sec:result_discussion}) when the use of video IDs is replaced with content embeddings. A recent study~\cite{ni2023content} has demonstrated the effectiveness of video encoders that use end-to-end training (VideoRec) to replace video ID in recommendation models for short videos. However, this approach comes with 10-50x computational cost over the ID baseline.

We propose a new framework of adapting content embeddings in ranking models with the flexibility of controlling generalization and memorization. Our method is based on item Semantic IDs (SIDs) which are originally proposed in TIGER \cite{rajput2023recommender} as a hierarchical, sequential and compact representation for generative retrieval. The hierarchical nature of SID offer the flexibilty of granuality control by using various levels of prefixes, and the sequential property draws the connection to subword tokenization, e.g., SentencePiece model (SPM) \cite{kudo2018subword} in LLMs. Notably, TIGER~\cite{rajput2023recommender} uses SIDs for generative retrieval where efficiency is not a primary consideration, while our work focuses on using Semantic IDs in resource-constrained and latency-sensitive production-scale ranking models, where the hashing and adaptation through embeddings is the key.

The detailed contributions are: (1) We propose two ways of adapting SIDs in recommendation models as a replacement of item IDs: n-gram and SPM. For both of them, the key idea is to create content-based hashing through sub-pieces of item SIDs, while SPM provides a learnable approach from item distribution by grouping sub-pieces with variable lengths; (2) We conduct extensive experiments on the YouTube dataset to demonstrate the effectiveness of our approaches. To that end, we show that SID-based adaption outperforms the directly using content embeddings. We also demonstrate the superior performance of SPM over n-gram when using large embedding tables with the same number of embedding lookups per item; (3) We also demonstrate the productionization of SIDs for a corpus of billions of videos in YouTube with examples of meaningful and granular hierarchical relationships, along with the success of replacing video IDs in the product scenario.

%% file: 2_related_work.tex
\paragraph{Embedding learning} Recommender models rely on learning good representation of categorical features. A common technique to encode categorical features is to train embeddings using one-hot embeddings. Word2vec~\cite{mikolov2013distributed} popularized this in the context of language models. Hashing trick~\cite{weinberger2009feature} is typically used when the cardinality is high, but it causes random collisions. Multiple hashing~\cite{zhang2020model} offers some relief but still leads to random collisions. Deep Hash Embedding ~\cite{kang2021learning} circumvents this problem by not maintaining embedding tables but at the cost of increased computation in the hidden layers. In contrast, we use Semantic IDs --- a compute-efficient way to avoid random collisions during embedding learning for item IDs. Semantic IDs improve generalization in recommender models by enabling collisions between semantically related items. 

\paragraph{Cold-start and content information} Content-based recommender models have been proposed to combat cold-start issues (e.g. ~\cite{schein2002methods}, ~\cite{volkovs2017content}) and to enable transferable recommendations (\cite{wang2022transrec}, ~\cite{hou2022learning}, ~\cite{ni2023content}). Recently, embeddings derived from content information are also popular (e.g., DropoutNet~\cite{volkovs2017dropoutnet}, CC-CC~\cite{shi2019adaptive} and ~\cite{du2020learn}). PinSage~\cite{Ying_2018} aggregates visual, text, and engagement information to represent items. Moreover, PinnerFormer~\cite{pancha2022pinnerformer} uses sequences of PinSage embeddings corresponding to item history to build a sequential recommendation model. In contrast to these efforts, our goal is to develop content-derived representations that not only generalize well. However, it can also improve performance relative to using item ID features which is a significantly challenging task. \cite{ni2023content} have successfully tackled the challenge of replacing video ID with content embedding derived from video encoders that are trained end-to-end with the recommendation model for short videos. In a similar vein, TransRec~\cite{wang2022transrec} also trains end-to-end and uses multiple modality information to represent items for enabling transferable recommendations. However, both approaches significantly increase training costs, making them challenging to deploy in production. Semantic IDs offer an efficient compression of content embeddings into discrete tokens, making it feasible to use content signals in production recommendation systems.
Furthermore, unlike PinnerFormer~\cite{pancha2022pinnerformer} which is used for offline inference, our focus is to improve generalization of a ranking model used for real-time inference.
Therefore, approaches that significantly increase resource costs (including storage, training and serving) make them infeasible to deploy in production. Semantic IDs offer an efficient compression of content embeddings into discrete tokens, making it feasible to use content signals in production recommendation systems. \cite{ni2023content} introduce a large dataset of short videos and show that existing video encoders do not produce embeddings that are useful for recommendations purposes.

\paragraph{Discrete representations} Several techniques exist to discretize embeddings, including VQ-VAE~\cite{van2017neural}, VQ-GAN~\cite{esser2021taming} and their variants used for generative modeling (e.g., Parti~\cite{yu2022scaling} and SoundStream~\cite{zegh2021soundstream}). TIGER~\cite{rajput2023recommender} used RQ-VAE in the context of recommender applications. Conventional techniques like Product Quantization~\cite{jegou2010product} and its variants are used by many recommender models (e.g., MGQE~\cite{kang2020learning} and ~\cite{hou2022learning}). However, these do not offer hierarchical semantics, which we leverage in our work.

%% file: 4_proposed_approach.tex
\subsection{Overview}
\label{subsec_overview}
Given content embeddings for a corpus of items, in contrast with the approach of directly using the embeddings as input feature, we propose an \textit{efficient} two-stage approach to leverage content signal in downstream recommendation models.

\begin{itemize}
    \item \textit{Stage 1: Efficient compression of content embeddings into discrete Semantic IDs}. We propose a Residual Quantization technique called RQ-VAE~\cite{rajput2023recommender, lee2022autoregressive, zegh2021soundstream} to quantize dense content embeddings into discrete tokens to capture semantic information about videos. This compression is crucial to allow us to efficiently represent a user's past history because each item can be efficiently be represented as a few integers rather than high-dim embeddings. Once trained, we freeze the trained RQ-VAE model and use it for training the downstream ranking model in Stage 2.

    \item \textit{Stage 2: Training the ranking model with Semantic IDs}. We use the model from Stage 1 to map each item to its Semantic ID and then train embeddings for Semantic ID, along with the rest of the ranking model (Section~\ref{sec:sid-vid-repr}). In practical scenario, ranking models are typically trained sequentially on recently logged data.
    
\end{itemize}

A key design choice in our proposal is to train and then freeze the RQ-VAE model from Stage 1. The frozen RQ-VAE model generates Semantic IDs for training and serving the ranking model. Recent data may include items that may not exist in the training distribution of the RQ-VAE model. This raises a potential concern from freezing the model, which could hurt the performance of the ranking model over time. As detailed in Appendix~\ref{sec:stability}, our analysis of YouTube ranking models utilizing Semantic IDs derived from RQ-VAE models trained on both older and recent data reveals comparable performance, indicating the stability of learned semantic representations over time.
% In our work, we use RQ-VAE~\cite{zegh2021soundstream} instead of VQ-VAE~\cite{van2017neural} to quantize content embeddings (Section~\ref{sec:rqvae}), since RQ-VAE organically captures hierarchical semantic structure (Figure~\ref{fig:combined-clusters}), offering interpretability.

\subsection{RQ-VAE for Semantic IDs (SIDs)}\label{sec:rqvae}
\begin{figure}[t]
\centering
\includegraphics[width=0.65\textwidth]{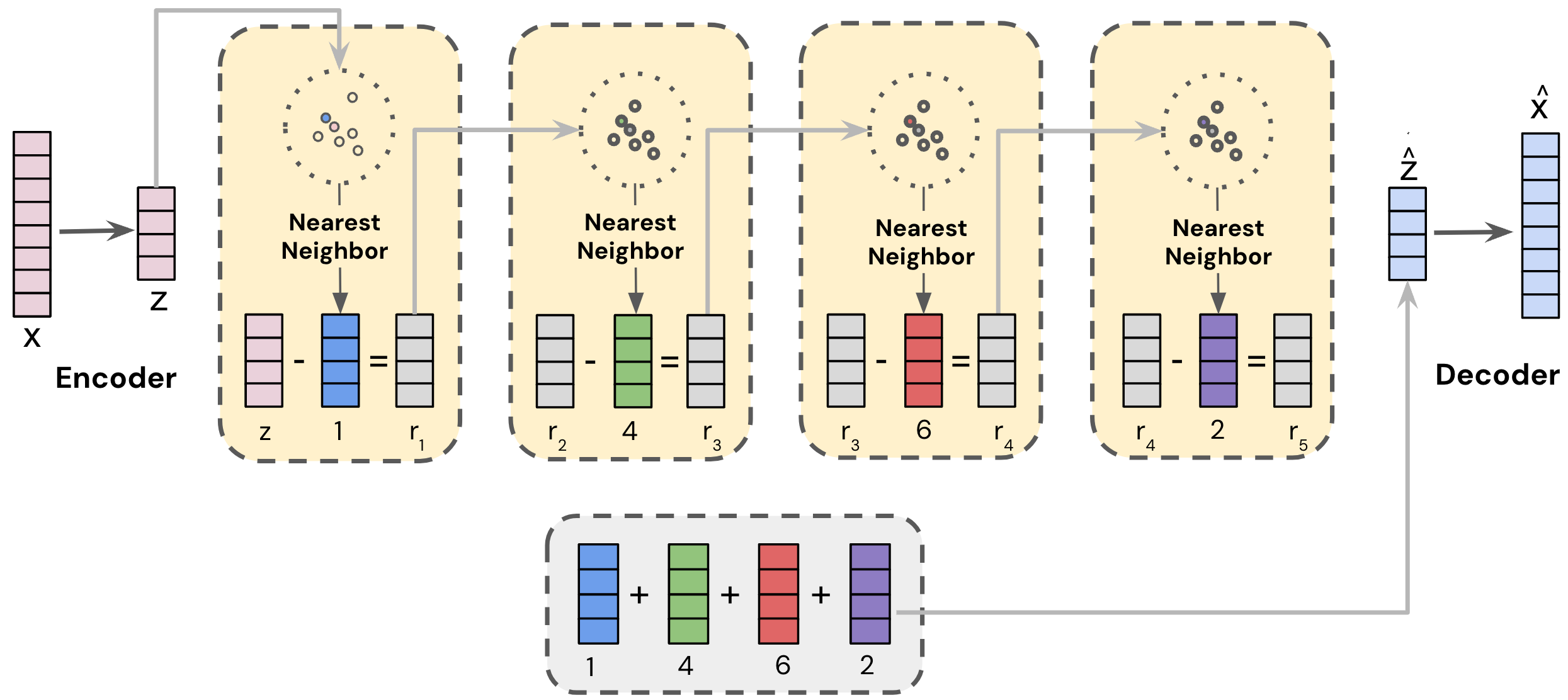}
\caption{Illustration of RQ-VAE: The input vector $\vx$ is encoded into a latent $\vz$, which is then recursively quantized by looking up the nearest codebook vector of the residual at each level. In this figure, the item represented by $\vx$ has $(1,4,6,2)$ as its Semantic ID.}
\vspace{-2.0em}
\label{fig:rq-vae}
\end{figure}
SIDs are generated from item content embeddings using Residual-Quantized Variational AutoEncoder (RQ-VAE)~\cite{lee2022autoregressive, zegh2021soundstream, rajput2023recommender} that applies quantization on residuals at multiple levels as shown in Figure~\ref{fig:rq-vae}. There are three jointly-trained components: (1) an encoder $\cE$ that maps the content embedding $\vx \in \mathbb{R}^D$ to a latent vector $\vz \in \mathbb{R}^{D'}$, (2) a residual quantizer with $L$ levels, each with a codebook $\cC_l := \set{\ve^{l}_k}_{k=1}^K$, where $\ve^{l}_k \in \mathbb{R}^{D'}$ and $K$ is the codebook size; the quantizer recursively quantizes the residual $\vr_l$ at each level $l$ to the nearest codebook vector $\ve_{c_{l}}$, and (3) a decoder $\cD$ that maps the quantized latent $\hat{\vz}$ back to the original embedding space $\hat{x}$. We use the following loss to train the RQ-VAE model:
$\cL  = \cL_{recon} + \cL_{rqvae}$, where $\cL_{recon} = \|\vx-\hat{\vx}\|^2$ and $\cL_{rqvae} = \sum_{l=1}^L\ \beta\|\vr_l-\text{sg}[\ve_{c_l}]\|^2 + \|\text{sg}[\vr_l] - \ve_{c_l}\|^2$ and sg denotes the stop-gradient operator. $\cL_{recon}$ aims to reconstruct the content embedding $\vx$. The first and the second terms in $\cL_{rqvae}$ encourages the encoder and the codebook vectors to be trained such that $\vr_l$ and $\ve_{c_l}$ move towards each other.

%Nikhil: Moved the hierarchy of concepts section to the Appendix.
% \subsubsection{Semantic IDs as hierarchy of concepts}
% We illustrate the hierarchy of concepts captured by Semantic IDs from the videos in our corpus. Section~\ref{sec:exp-setup} details the hyper-parameters used to train the RQ-VAE model. Intuitively, we can think of Semantic IDs as forming a trie over videos, with higher levels representing coarser concepts and lower levels representing more fine-grained concepts. Figure~\ref{fig:sports-cluster} shows an example sub-trie from our trained RQ-VAE model that captures a hierarchy of concepts within sports using 4 tokens. We also show the average pairwise cosine similarity in the content embedding space for videos with a shared Semantic ID prefix in Appendix~\ref{sec:cosine_similarity_analysis}.

% \begin{figure}[!t]
% \centering
% \includegraphics[width=0.4\textwidth]{sports-cluster-2.png}
% \caption{A sub-trie that exhibits the hierarchical structure for sports videos within Semantic IDs.}
% \label{fig:sports-cluster}
% \end{figure}

\subsection{Semantic ID Representation in Ranking}\label{sec:sid-vid-repr}

In this section, we discuss how we model item representations derived from SIDs to use in ranking models. For a given item $v$, an RQ-VAE model with $L$ levels generates a SID as a sequence $(c^{v}_1, ... c^{v}_L)$. The idea of adaptation is to create subwords for hashing the SID sequence into a number of learnable embeddings. We propose two techniques for the adaptation:\\

\noindent \textbf{N-gram-based:} N-gram item representations leverage SID codes by grouping them into subwords of length N. Each subword is then associated with a learnable embedding, effectively capturing the semantic relationships within the N-gram. The item representation is constructed by summing the embeddings of all N-gram subwords in an item. For instance, a unigram representation would have L subwords, each containing a single code: ${(c^{v}_1), ..., (c^{v}_L)}$. A bigram representation with non-overlapping codes would consist of L/2 subwords, each containing two consecutive codes: ${(c^{v}_1,c^{v}2), ..., (c^{v}_{L-1}, c^{v}_L)}$. To associate learnable embeddings with these N-gram-based subwords, a separate embedding table is learned for each subgroup. Since each code has a cardinality of K, the embedding table for an N-gram group contains $K^N$ rows. These embedding tables are jointly trained with the other parameters of the ranking model, enabling the network to learn representations that effectively capture the relationship between semantic codes within the context of the ranking task.\\

\noindent\textbf{SPM-based:} While N-gram-based video representations offer a straightforward approach to capture relationships between sequential codes in Semantic ID, they suffer from limitations that hinder their effectiveness. First, their reliance on fixed grouping based on predefined N-gram sizes restricts their ability to adapt to the specific characteristics of the Semantic ID corpus, leading to suboptimal embedding table lookups. Second, the number of rows in the embedding tables in N-gram grow exponentially with N, imposing a significant memory burden. These challenges motivate adaptation of Semantic IDs with Sentence Piece Models (SPM) \cite{kudo2018subword}, which offer a more adaptive and efficient solution for representing item content. We propose using SPM to dynamically learn Semantic ID subwords based on the distribution of impressed items. This allows dynamic length subwords such that popular co-occuring codes are automatically comined as a single subgroup, whereas codes that rarely co-occur together may fallback to unigram. For SPM-based representation, we learn a single embedding table where each row corresponds to a particular variable-length subpieces. By adaptively constructing subword vocabularies given a fixed embedding table size, the SPM vocabulary allows striking a balance between generalization and memorization.\\

%% file: 5_experiments.tex
% We evaluate the performance from using Semantic ID-based video representation for the ranking task.
\subsection{Experimental Setup}{\label{sec:exp-setup}}
\noindent \textbf{Ranking Model.} We conduct our experiments on a multitask production ranking model \cite{Tang_2023, mtlranking}, which is used for recommending the next video to watch, given a video a user is watching and user's past activities. This model uses O(10) million buckets for random hashing to accommodate O(100) millions of videos in our corpus and is trained sequentially on logged data. In the baseline, random hashing of video IDs is used for three key features: users' watch history, watch video, and the candidate video to be ranked. We evaluate our methods on the data that the trained model has not yet seen, allowing us to understand the performance under the data-distribution shift of the video corpus. 

The inherent scale and real-time demands of ranking models necessitate embedding tables with specific characteristics to ensure efficient and effective performance. Firstly, the embedding table size needs to easily fit in the memory. This was one of our key considerations when deciding N in the N-gram-based Semantic ID representations. Since the number of rows in the embedding tables grow exponentially with N, we limit our analysis to $N \leq 2$ for N-gram-based representations. Secondly, the embedding lookups need to be fast to provide near-instantaneous responses to user requests. Our analysis is grounded in the above two properties.\\

\noindent \textbf{Content Embeddings.} Semantic IDs are generated using dense content embeddings. We use a video encoder to generate dense content embeddings for each YouTube video. The video encoder is a transformer model that uses Video- BERT~\cite{videoBert} as the backbone architecture, takes audio and visual features as inputs, and outputs $2048$-dimensional embeddings that capture the topicality of the video. This model was trained using techniques described in ~\cite{Lee_2020_CVPR}.\\

\noindent \textbf{Experimental Settings.} We compare the two proposed Semantic ID-based representations with two baseline representation techniques: directly using raw content embeddings referred to as \textit{Dense Input}, and the commonly used randomized hashed IDs referred to as \textit{Random Hashing}. Since directly using dense input embeddings as item representation obviate the need for embedding table parameters, we also introduce additional baselines for the Dense Input approach for a fair comparison, where we increase the ranking model layers by 1.5x and 2x to study how increasing the model depth affects the ranking performance. To generate the Semantic IDs, we use $L=8$ depth resulting in 8 codes in the Semantic ID of each video. The codebook size for RQ-VAE was set to $K=2048$.\\

\noindent \textbf{Evaluation metrics} The ranking model is trained sequentially on the first $N$ days of data, where each day contains logged data generated from user interactions on that day. We evaluate the model's performance using AUC for CTR for the data from ($N+1$)-th day. We further slice the metric on items introduced on the ($N+1$)-th day. We refer to this as CTR/1D.
CTR AUC and CTR/1D AUC metrics evaluate the model's ability to generalize over time due to data distribution shifts and cold-start items, respectively. A $0.1\%$ change in CTR AUC is considered significant for our ranking model.
\subsection{Performance of Semantic ID}\label{sec:result_discussion}
\begin{figure}[!ht]
\begin{subfigure}{0.495\linewidth} 
    \centering
    \includegraphics[width=\linewidth]{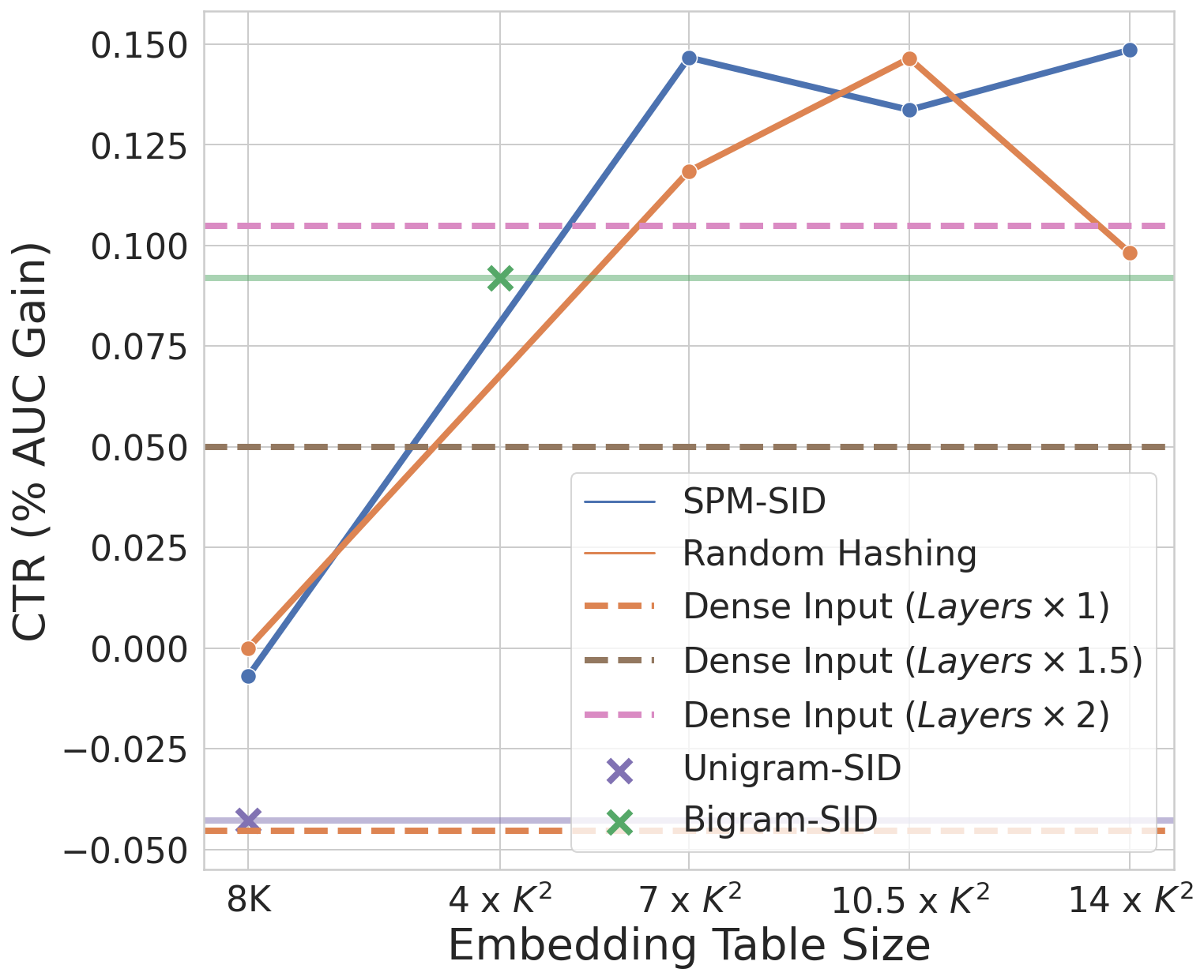}
    \caption{Overall CTR AUC}
    \label{fig:ctr_noum}
\end{subfigure}
\hfill
\begin{subfigure}{0.495\linewidth} 
    \centering
    \includegraphics[width=\linewidth]{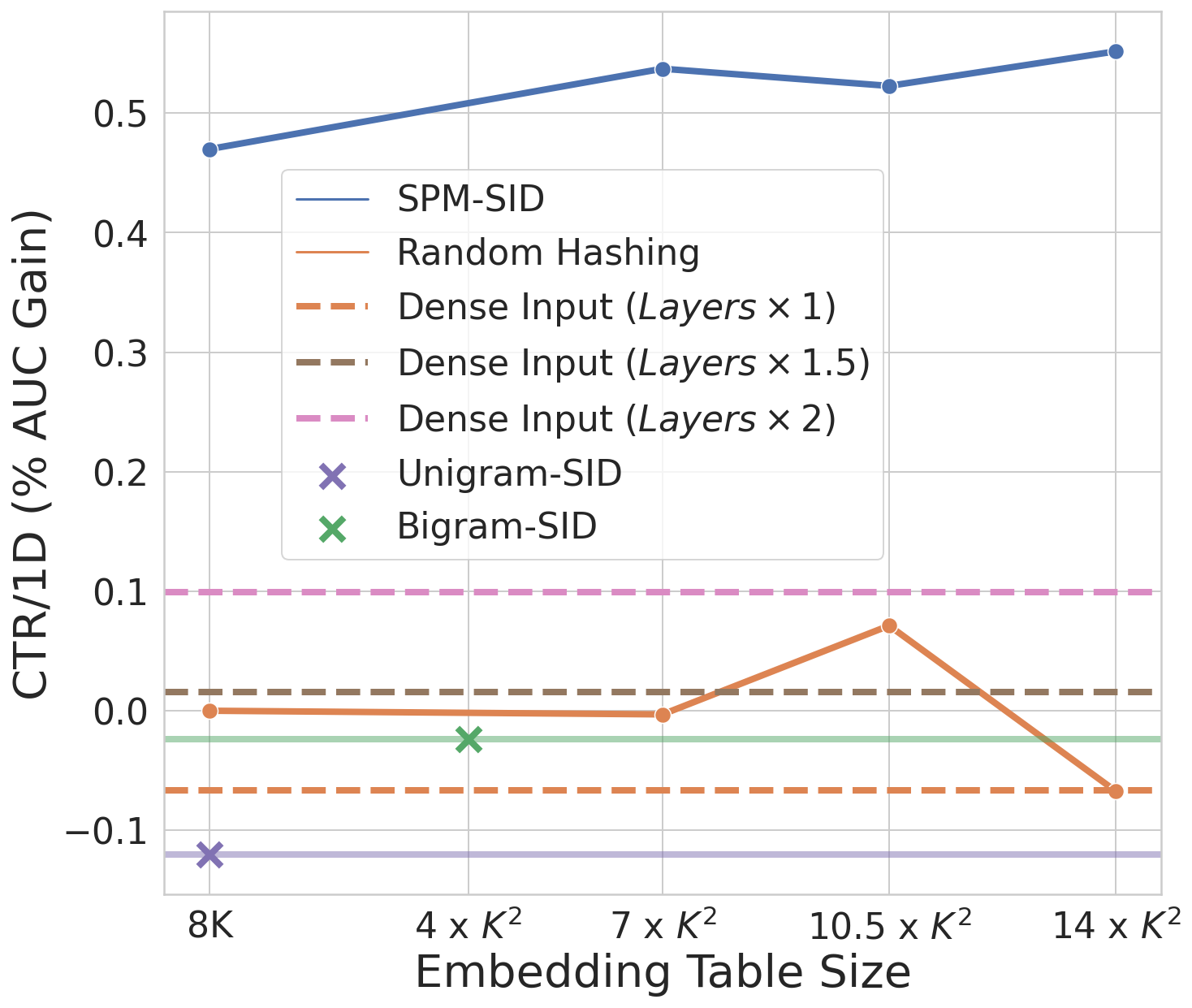}
    \caption{Cold-start CTR/1D AUC}
    \label{fig:ctr_1d_noum}
\end{subfigure}
\caption{Percentage improvement in CTR AUC metric when user history is not used as a input feature. Improvement is relative to Random Hashing baseline with 8K embedding table size.}
\label{fig:results_without_wh}
\vspace{-1em}
\end{figure}

\begin{figure}[!t]
\centering
\begin{subfigure}{0.49\linewidth} 
    \centering
    \includegraphics[width=\linewidth]{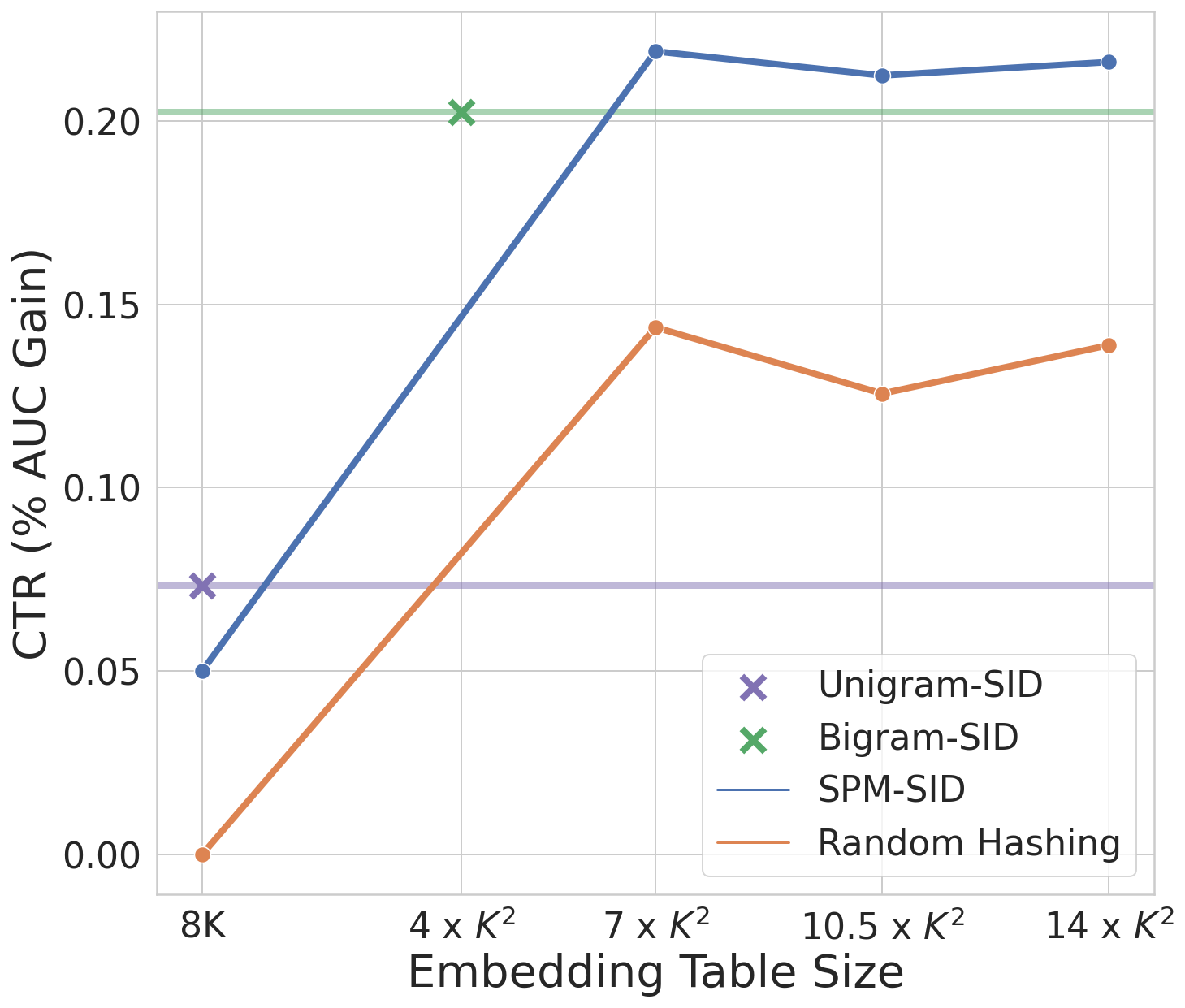}
    \caption{Overall CTR AUC}
    \label{fig:ctr_um}
\end{subfigure}
\hfill
\begin{subfigure}{0.49\linewidth} 
    \centering
    \includegraphics[width=\linewidth]{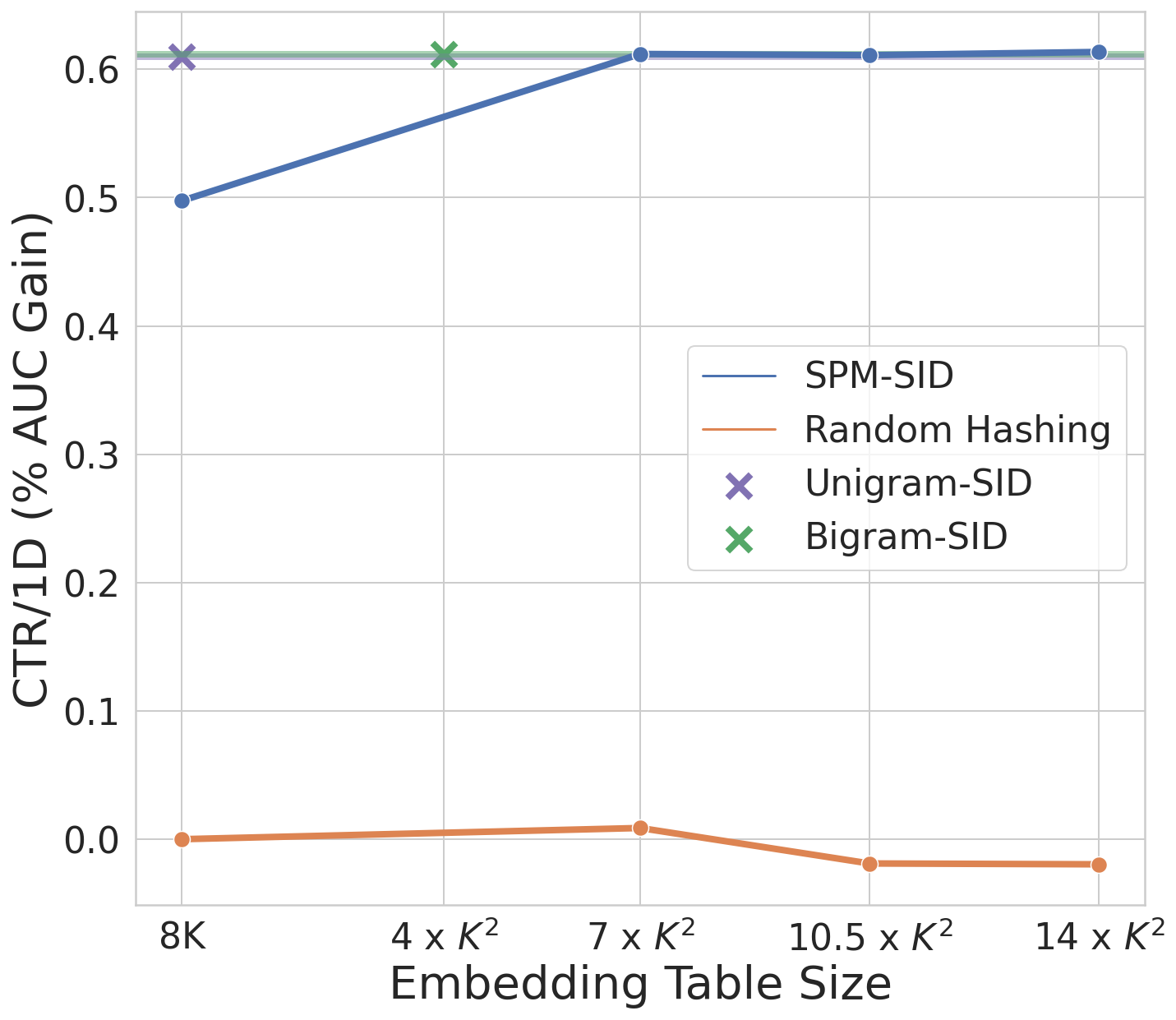}
    \caption{Cold-start CTR/1D AUC}
    \label{fig:ctr_1d_um}
\end{subfigure}
\caption{Percentage improvement in CTR AUC metric when user history is used as a input feature. Improvement is relative to Random Hashing baseline with 8K embedding table size.}
\label{fig:results_with_wh}
\vspace{-1em}

\end{figure}

Storing content embeddings for each video in users' watch history is highly resource intensive. Hence, training a baseline large-scale ranking model that uses content embeddings to represent each video in users' watch history is infeasible. To better understand which representation method performs better, we consider two settings of the ranking model. First, we compare the SID-based representation with raw content embeddings and random hashing based ID such that user history is not used as an input feature (Figure \ref{fig:results_without_wh}). In this setting, two video features (i.e., current and candidate video) are used as input features to the ranking model. In the second setting, we use users' watch history as the input feature (along with current and candidate video), where the SID-based representation is compared with random hashing (Figure \ref{fig:results_with_wh}).\\

\noindent\textbf{Dense Content Embedding vs. Random Hashing.} We observe that directly using content embeddings (Dense Input) to replace random hashing-based IDs, without additional changes to the model architecture, doesn't lead to better quality. As shown in figures~\ref{fig:ctr_noum}-\ref{fig:ctr_1d_noum}, the Dense Input baseline performs worse than the video-ID based baseline. We hypothesize that that the ranking models heavily rely on memorization from the ID-based embedding tables; replacing the embedding table with fixed dense content embeddings as a feature leads to poor CTR. For testing this hypothesis, we also ran experiments with 1.5x-2x layers in the ranking model to increase the model's memorization ability for the Dense Input baseline. We found that increasing the depth does improve quality for both overall and cold-start items compared to the random hashing-baseline. In fact, the increase in CTR is higher for the Dense Input Model with 2x layers compared to Dense Input with 1.5x layers, indicating more the number of layers, better the memorization (Overall CTR) and generalization (cold-start CTR/1D). However, increasing the number of layers can cause the serving cost to increase considerably. As discussed below, SIDs allows retaining the semantic information from raw content embeddings, while still flexibly and efficiently providing memorization via learned embedding tables.

\begin{figure}[!t]
\centering
\includegraphics[width=0.6\textwidth]{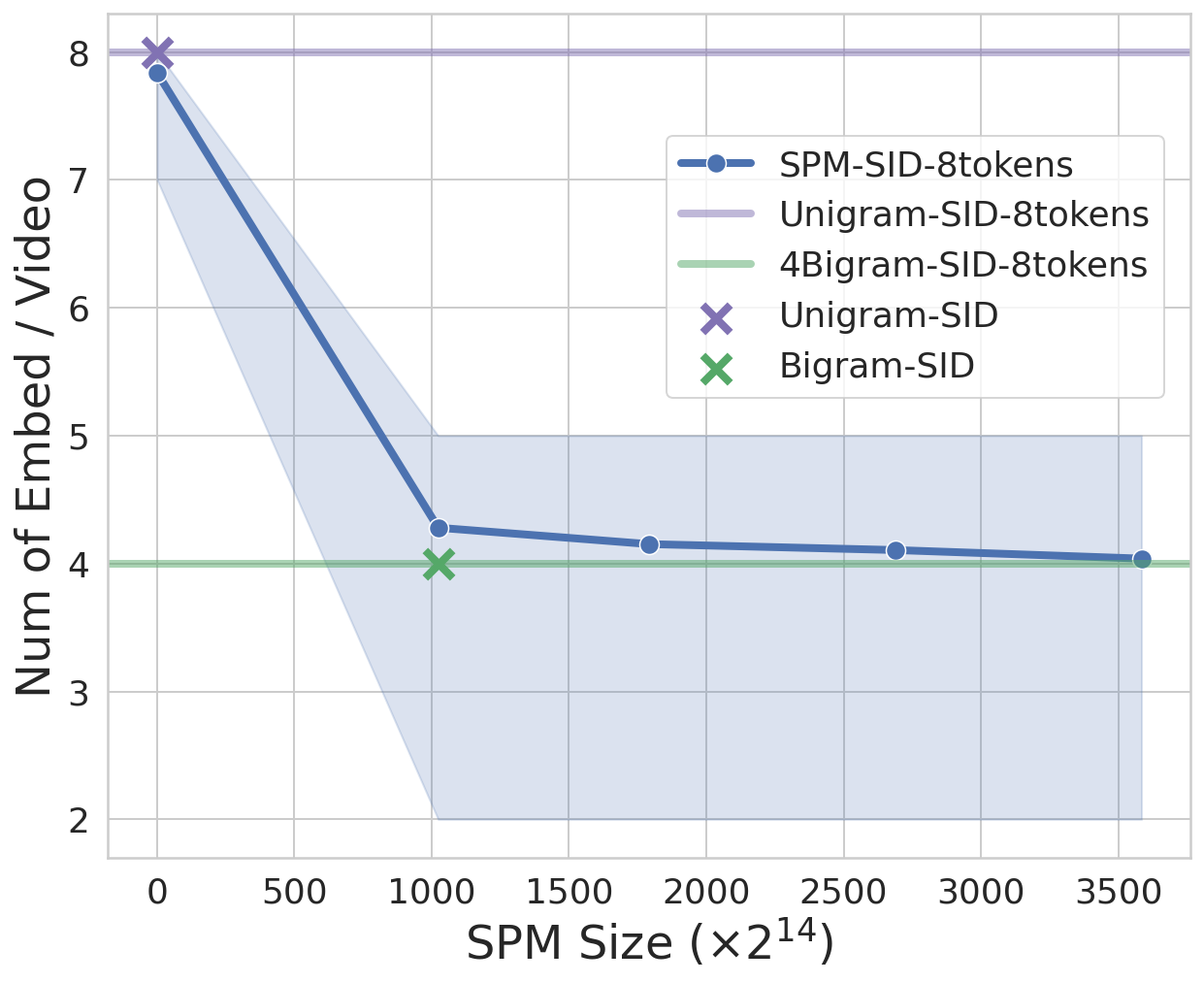}
\caption{Number of Subword Embeddings per video.}
\label{fig:num_embeddings}
\vspace{-2em}
\end{figure}

\noindent\textbf{SID vs. Baselines.} We compare the two types of SID representations (N-gram and SPM) with the baselines, where for N-gram-SID, we use Unigram (N=1) and Bigram (N=2). When using N-gram, the embedding table size is based on all the possible combinations for the respective N-gram, i.e., Unigram-SID has $8 \times K$ rows and Bigram-SID has $4 \times K^2$ rows, respectively. We found that both Unigram-SID and Bigram-SID lead to worse overall CTR compared to Random Hashing when the user history is not used as an input feature (Figure~\ref{fig:results_without_wh}). This could be because of skew in the content in the training data, causing sparse usage of the embedding table. This issue doesn't occur in random hashing since the embeddings are uniformly used due to random assignment of videos to embeddings in the embedding table. On the other hand, when we use the user history as an input feature (Figure~\ref{fig:results_with_wh}), both Unigram-SID and Bigram-SID perform much better than random hashing because the video content in users' watch history is likely covering more diverse content, leading to more uniform usage of the embedding table. Next, we show impressive gains from the SPM-SID-based video representations. While SPM-SID consistently outperformed N-gram representations when employing larger embedding tables, particularly evident in the improved CTR/1D AUC metrics (see Figures ~\ref{fig:ctr_1d_noum} and ~\ref{fig:ctr_1d_um}), suggesting greater generalization capabilities towards cold-start items, a nuanced observation emerges for smaller embedding table sizes. Specifically, when the embedding table size is limited ($8 \times K$ or $4 \times K^2$), N-gram methods demonstrate a slight advantage over SPM-SID. This behavior can be attributed to the smaller subword vocabulary learned by SPM within these constrained table sizes, potentially hindering its ability to fully capture complex semantic relationships. Note that for most production ranking models, a large embedding table is necessary for good quality. Hence, the SPM-SID based representation is more beneficial for large-scale production ranking models. Overall, both Bigram-SID and SPM-SID significantly outperformed random hashing in our experiments with large-scale ranking models, highlighting the importance of structured representations for capturing semantic relationships in improving cold-start video recommendations.\\

\noindent\textbf{Efficiency in SPM-SID vs. N-gram-SID.} 
In contrast to N-gramSID representations, which utilize fixed embedding table sizes, SPM-SID offers the flexibility of adapting to a given embedding table size. This adaptation is achieved through the construction of subwords directly based on the training data. Given a fixed embedding table, SPM dynamically generates subwords, each mapping to a unique table entry. This optimizes Semantic ID representation within the size constraint, improving video representation efficiency. Moreover, in terms of embedding table lookups SPM-SID is more optimal compared to N-gram-SID. We plot the number of embedding lookups per video vs. the embedding table size in figure~\ref{fig:num_embeddings}. The plot highlights the adaptive nature of SPM, where the number of lookups are dynamically reduced for the head/common videos in the training data, while the average number of lookups are comparable to the fixed number of lookups in N-gram. This adaptive nature of SPM contributes to its enhanced efficiency and scalability, making it a more suitable approach for large-scale ranking models.

%% file: 6_conclusion.tex
This paper tackles the challenging task of removing reliance on widely used item IDs in recommendation models. Using the YouTube ranking model as a case study, we discuss the disadvantages of using item ID features in large-scale production recommendation models. Using RQ-VAE, we develop Semantic IDs for billions of YouTube videos from frozen content embeddings to capture semantically meaningful hierarchical structures across the corpus. We propose and demonstrate Semantic IDs as an effective method for replacing video IDs to improve generalization by introducing meaningful collisions. %In contrast to content embedding as a feature, Semantic IDs offer a compact content representation for videos that makes it feasible to use content signals for user watch history features, which are critical for personalization. Moreover, our proposed approach is computationally efficient and does not significantly increase training time for the ranking model.
% As for future work, we plan to investigate the generalization benefits of Semantic IDs with varying levels and codebook sizes. Beyond ranking models, we are actively exploring the application of Semantic IDs in two-tower retrieval models and Transformer-based models for sequential recommendations, yielding early promising improvements in model generalization.

%% file: appendix.tex
\section{Appendix}
\label{sec:appendix}

\subsection{RQ-VAE Training and Serving Setup}
\label{sec:rqvae_hparams}

\textbf{Model Hyperparameters.} For the RQ-VAE model, we use a $1$-layer encoder decoder model with dimension 256. We apply $L=8$ levels of quantization using codebook size $K=2048$ for each.

\noindent \textbf{RQ-VAE Training}: We train the RQ-VAE model on a random sample of impressed videos until the reconstruction loss stabilizes ($\approx$10s of millions of steps for our corpus). Vector quantization techniques are known to suffer from \textit{codebook collapse}~\cite{dhariwal2020jukebox} during training, where the model only uses a small proportion of codebook vectors. To address this challenge, we reset unused codebook vectors at each training step to content embeddings of randomly sampled videos from within the batch~\cite{zegh2021soundstream}, which significantly improved the codebook utilization. We use $\beta=0.25$ to compute the training loss. Once trained, we freeze the RQ-VAE model and use the encoder to produce Semantic IDs for videos.

\noindent \textbf{RQ-VAE Serving/Inference}: As new videos get introduced into the corpus, we generate the Semantic IDs using the \emph{frozen} RQ-VAE model. Semantic IDs are then stored and served similarly to other features used for ranking.

\subsection{Stability of Semantic IDs over time}\label{sec:stability}
\begin{figure}[!h]
    \centering
    \begin{subfigure}[b]{0.49\linewidth}
        \centering
        \includegraphics[width=\linewidth]{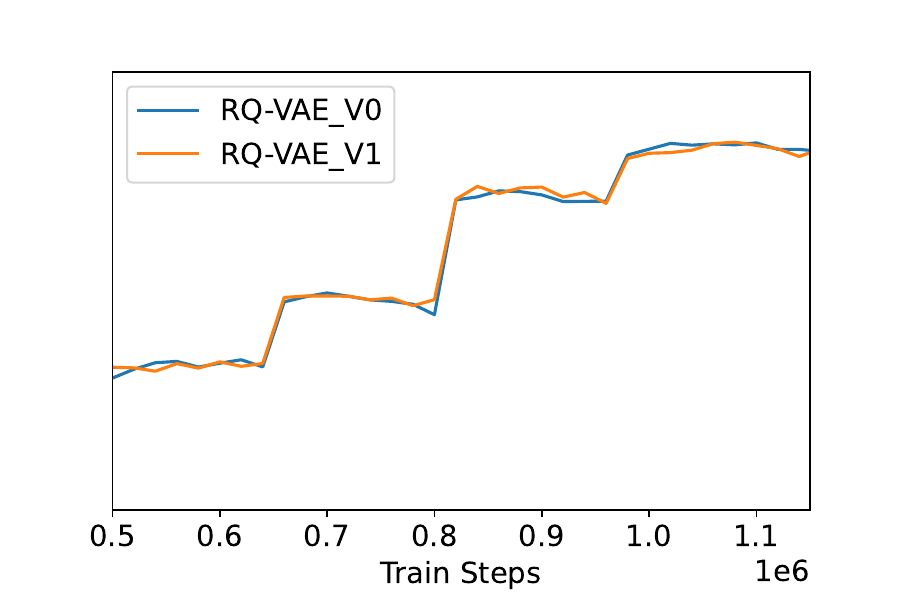} 
        \caption{CTR/AUC}
        \label{fig:rq-vae-stability_ctr}
    \end{subfigure}
    \begin{subfigure}[b]{0.49\linewidth}
        \centering
        \includegraphics[width=\linewidth]{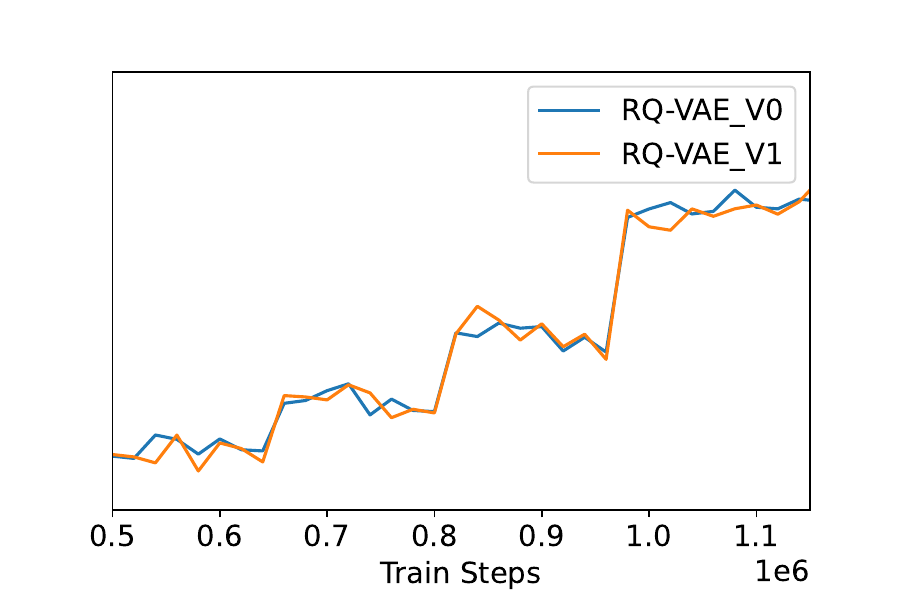}
        \caption{CTR-1D/AUC}
        \label{fig:rq-vae-stability_ctr1d}
    \end{subfigure}
    \caption{Comparison of ranking performance of SID-3Bigram-sum representation derived using RQ-VAE$_{v0}$ and RQ-VAE$_{v1}$.}
    \label{fig:rq-vae-stability}
\end{figure}

To study Semantic IDs' stability, we train two RQ-VAE models: RQ-VAE$_{v0}$ and RQ-VAE$_{v1}$, using data ~6 months apart. Figure~\ref{fig:rq-vae-stability} shows that the performance of the production ranking model trained on recent engagement data (using SID-3Bigram-sum) are comparable for Semantic IDs derived from both RQ-VAE$_{v0}$ and RQ-VAE$_{v1}$. This confirms that semantic token space for videos learned via RQ-VAE is stable for use in the downstream production ranking model over time.

\subsection{Semantic IDs as hierarchy of concepts}

\begin{figure*}[!ht]
\centering
\includegraphics[trim = 2cm 2cm 1cm 3cm, clip, width=0.9\textwidth]{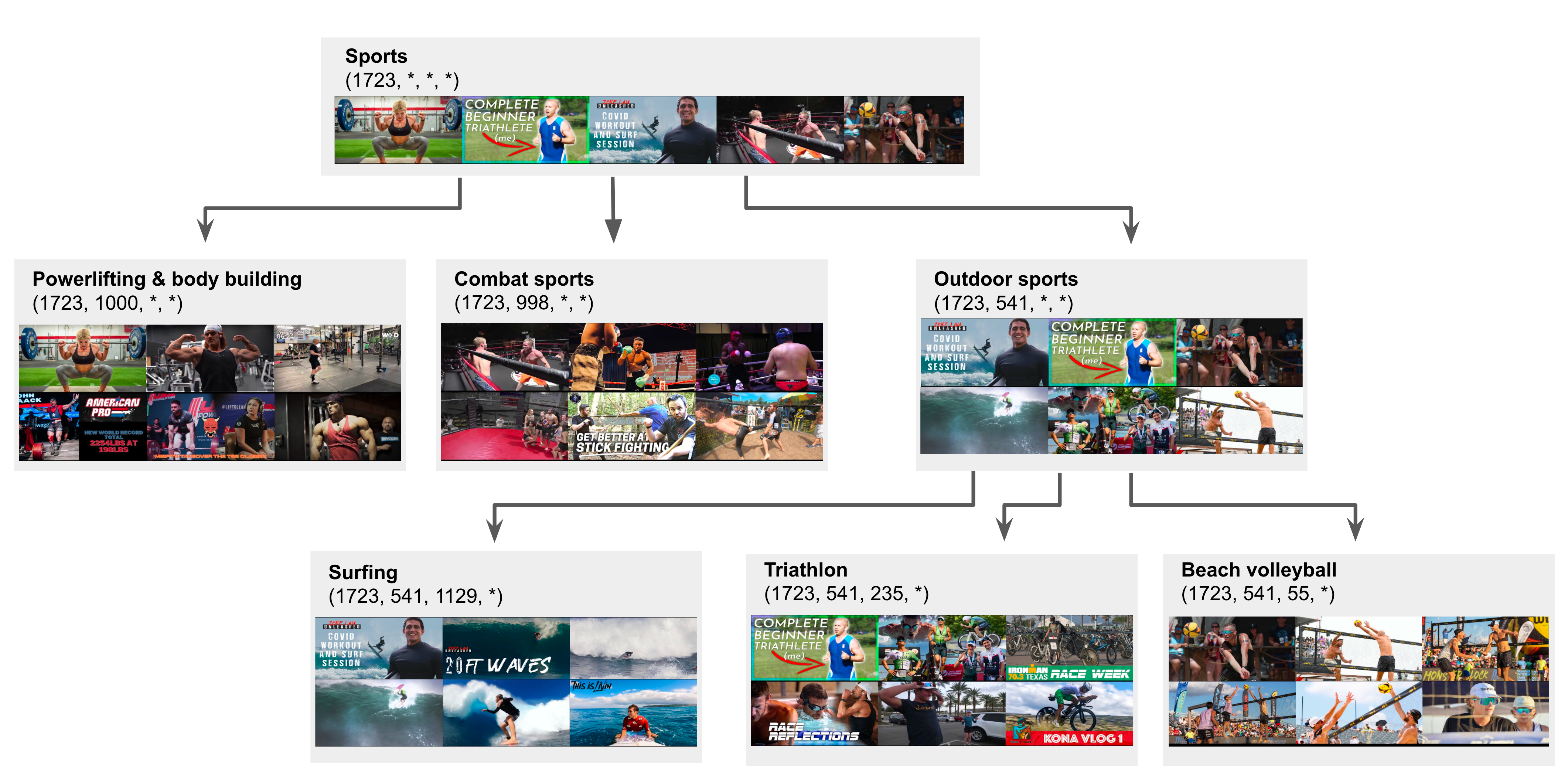}
\caption{A sub-trie that exhibits hierarchical structures with Semantic IDs capturing sports videos.}
\label{fig:sports-cluster}
\end{figure*}

\begin{figure*}[!ht]
\centering
   \includegraphics[trim = 3cm 4cm 4cm 1.5cm, clip, width=0.9\textwidth]{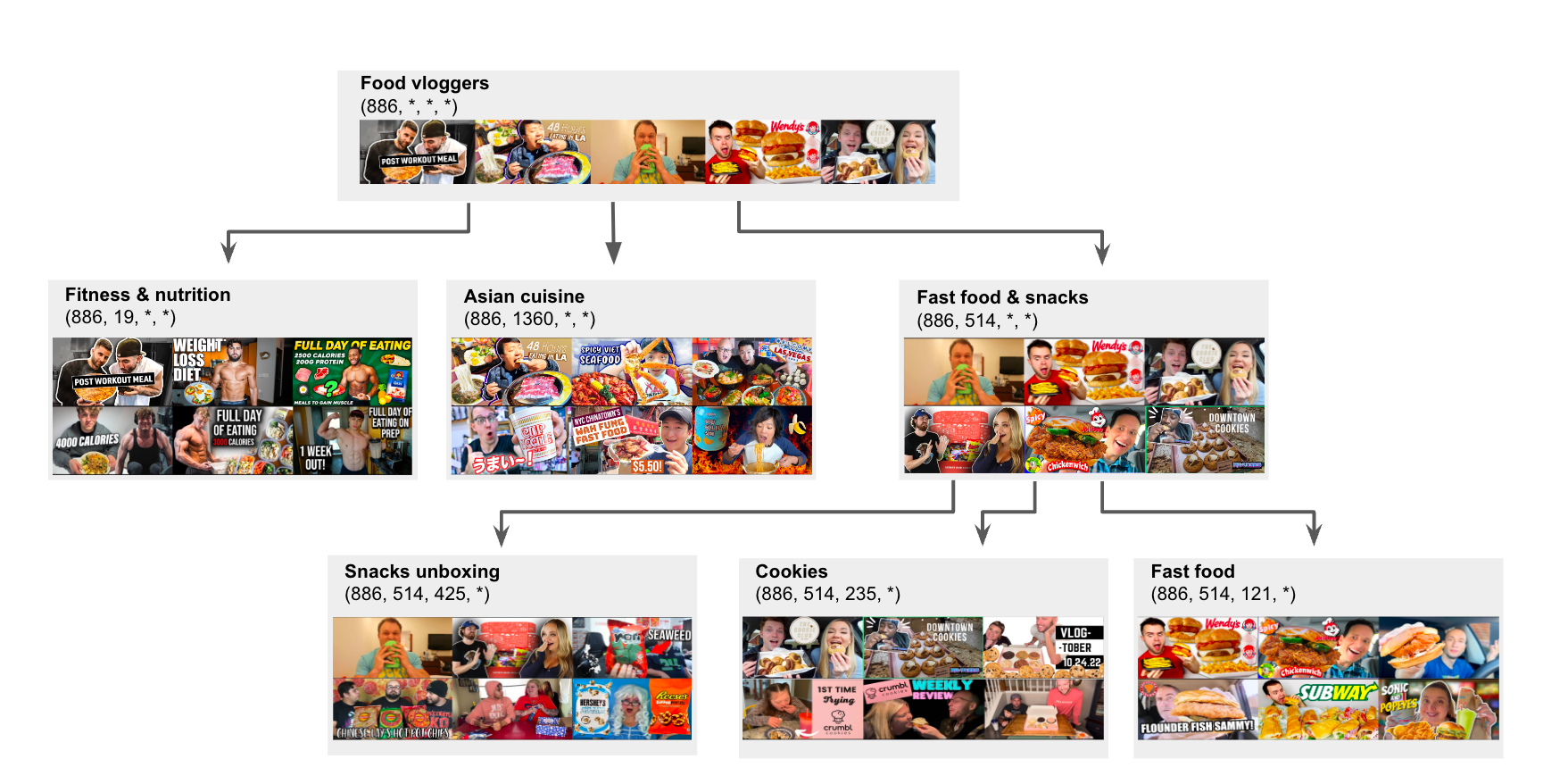}
\caption{A sub-trie that exhibits hierarchical structures with Semantic IDs capturing food vlogging videos.}
\label{fig:food-cluster}
\end{figure*}

We illustrate the hierarchy of concepts captured by Semantic IDs from the videos in our corpus. Section~\ref{sec:exp-setup} details the hyper-parameters used to train the RQ-VAE model. Intuitively, we can think of Semantic IDs as forming a trie over videos, with higher levels representing coarser concepts and lower levels representing more fine-grained concepts. Figures \ref{fig:sports-cluster} and \ref{fig:food-cluster} show two example sub-tries from our trained RQ-VAE model with 4 tokens that captures a hierarchy of concepts within sports and food vlogging videos. 
% We illustrate the hierarchy of concepts captured by Semantic IDs from the videos in our corpus. Section~\ref{sec:exp-setup} details the hyper-parameters used to train the RQ-VAE model. Intuitively, we can think of Semantic IDs as forming a trie over videos, with higher levels representing coarser concepts and lower levels representing more fine-grained concepts. Figure~\ref{fig:combined-clusters} shows two example sub-tries from our trained RQ-VAE model that captures a hierarchy of concepts within sports using 4 tokens. We also show the average pairwise cosine similarity in the content embedding space for videos with a shared Semantic ID prefix in Appendix~\ref{sec:cosine_similarity_analysis}.
% \begin{figure}[!t]
% \centering
% \includegraphics[width=0.4\textwidth]{sports-cluster-2.png}
% \caption{A sub-trie that exhibits the hierarchical structure for sports videos within Semantic IDs.}
% \label{fig:sports-cluster}
% \end{figure}

\subsection{Similarity Analysis with Semantic ID}
\label{sec:cosine_similarity_analysis}
Table \ref{tab:sem-id-aggregate} shows the average pairwise cosine similarity in the content embedding space for all videos with a shared Semantic ID prefix of length $n$ and their corresponding sub-trie sizes. We consider two videos with Semantic IDs $(1,2,3,4)$ and $(1,2,6,7)$ to have a shared prefix of length $2$. We observe that as the shared prefix length increases, average pairwise cosine similarity increases while the sub-trie size decreases. These suggest that Semantic ID prefixes represent increasingly granular concepts as their lengths increase.

\begin{table}[!h]
\centering
%$\begin{adjustbox}{width=0.9\linewidth}
\resizebox{\columnwidth}{!}{%
    \begin{tabular}{clllc}
    \toprule
    & Shared prefix length  & Average pairwise cosine similarity & Typical sub-trie size   \\
    \cmidrule{2-4}
    & 1 & 0.41 & 150,000-450,000 \\
    & 2 & 0.68 & 20-150 \\
    & 3 & 0.91 & 1-5 \\
    & 4 & 0.97 & 1 \\
    \bottomrule
    \end{tabular}
}
\caption{Aggregate metrics for videos sharing Semantic ID prefix of length $n$. The typical sub-trie size refers to the 25th-75th percentile range (with rounding).}
\label{tab:sem-id-aggregate}
\end{table}

% \begin{figure*}[!ht]
% \centering
% \includegraphics[width=0.9\textwidth]{random_bigram_SPM_illustration.png}
% \caption{Illustration of random hashing, Bigram-SID and SPM-SID.}
% \label{fig:illustration}
% \end{figure*}